\documentclass[aps,pre,final,twocolumn,groupedaddress,showpacs]{revtex4-1}

\usepackage{mathptmx}       % selects Times Roman as basic font
\usepackage{helvet}         % selects Helvetica as sans-serif font
\usepackage{courier}        % selects Courier as typewriter font
\usepackage{type1cm}        % activate if the above 3 fonts are
\usepackage{makeidx}         % allows index generation
\usepackage{graphicx}        % standard LaTeX graphics tool
\usepackage[bottom]{footmisc}% places footnotes at page bottom
\usepackage{textcomp}        %needed for \textonehalf
\usepackage{amsmath}
\usepackage{amssymb}
\usepackage{xcolor}
\usepackage{enumitem}

\newcommand{\trac}{{\rm Tr}}

\newcommand{\diag}{{\rm diag}}

\newcommand{\be}{\begin{equation}}
\newcommand{\ee}{\end{equation}}
\newcommand{\bea}{\begin{eqnarray}}
\newcommand{\eea}{\end{eqnarray}}
\newcommand{\lt}{\left(}
\newcommand{\rt}{\right)}
\newcommand{\lgr}{\left\{}
\newcommand{\rgr}{\right\}}

\newcommand{\onehalf}{{\mbox{\textonehalf\,}}}
\newcommand{\quot}[1]{{``#1''}}
\newcommand{\vect}{{\mbox{vect}}}

\begin{document}

\title{ Steepest Entropy Ascent Model for Far-Non-Equilibrium Thermodynamics.\\ Unified  Implementation of the  Maximum Entropy Production Principle}

%\classification{03.65.Ta, 11.10.Lm, 04.60.-m, 05.45.-a}

%\keywords {Nonlinear Dynamics; Irreversibility;  Entropy;
% Information Theory;  Quantum Thermodynamics}

\author{Gian Paolo Beretta}
\affiliation{
Universit\`a di Brescia, via Branze 38, 25123
Brescia, Italy}

\email{gianpaolo.beretta@unibs.it}

\date{\today}
\begin{abstract}
By suitable reformulations, we cast the mathematical frameworks of  several  well-known different approaches to the description of non-equilibrium dynamics into a unified formulation valid in all these contexts, which extends to such  frameworks  the concept of Steepest Entropy Ascent (SEA) dynamics introduced  by the present author in previous works on quantum thermodynamics. Actually, the present formulation constitutes a generalization also for the quantum thermodynamics framework. The  analysis emphasizes that in the SEA modeling principle a key role is played by the geometrical metric with respect to which to measure the length of a trajectory in state space.  In the near thermodynamic equilibrium limit,  the metric tensor turns is  directly related to the Onsager's generalized resistivity tensor.  Therefore, through the identification of a suitable  metric field which generalizes the Onsager generalized resistance to the arbitrarily far non-equilibrium domain,  most of the existing theories of non-equilibrium thermodynamics can be cast in such a way that the state exhibits the spontaneous tendency to evolve  in state space along the path of SEA  compatible with the conservation constraints and the boundary conditions.
The resulting unified family of  SEA dynamical models are all intrinsically and strongly consistent with the second law of thermodynamics. The nonnegativity of the  entropy production  is a general and readily proved feature of SEA dynamics. In  several of the different approaches  to non-equilibrium description we consider here, the SEA concept has not been investigated before. We believe it defines the precise meaning and the domain of general validity of the so-called Maximum Entropy Production principle. Therefore, it is hoped that the present unifying approach may prove useful in providing a fresh basis for effective, thermodynamically consistent, numerical models and theoretical treatments of irreversible conservative relaxation towards equilibrium from far non-equilibrium states. The  mathematical frameworks are: A) Statistical or Information Theoretic Models of Relaxation; B)  Small-Scale and Rarefied Gases Dynamics (i.e., kinetic models for the Boltzmann equation); C)  Rational Extended Thermodynamics, Macroscopic Non-Equilibrium Thermodynamics, and Chemical Kinetics; D) Mesoscopic Non-Equilibrium Thermodynamics, Continuum Mechanics with Fluctuations; E) Quantum Statistical Mechanics, Quantum Thermodynamics, Mesoscopic Non-Equilibrium Quantum Thermodynamics, and Intrinsic Quantum Thermodynamics.
\end{abstract}


\maketitle

\section{Introduction}

The problem of understanding entropy and irreversibility has been
tackled by a large number of preeminent scientists during the past
century. Schools of thought have formed and flourished around
different perspectives of the problem. Several modeling approaches have been developed in various frameworks to deal with the many facets of non-equilibrium.

In this paper, we show how to construct Steepest Entropy Ascent (SEA)  models of non-equilibrium dynamics by adopting a unified mathematical formulation that allows us to do it at once in several different well-known frameworks of non-equilibrium description.

To avoid doing inevitable injustices to the many pioneers of all these approaches and to the many and growing fields of their application, here we skip a generic introduction and given no references nor a review of previous work. Rather, we dig immediately into the mathematical reformulations of the different frameworks in such a way that then the construction of the proposed SEA dynamics becomes formally a single geometrical problem that can be treated at once.

 Our reformulations here not only  allow  a precise meaning, general implementation, and unified treatment of the so-called Maximum Entropy Production  (MEP) principle (for a recent review see \cite{Martyushev}) in the various frameworks, but also extends to all frameworks an observation that we have been developing in the quantum thermodynamics framework for the last three decades \cite{Frontiers,LectureNotes,Gheorghiu,ROMP}. In doing so, we introduce an important generalization also for the quantum thermodynamics modeling framework.

 The observation is that we cannot simply  maximize the entropy production  subject to a set of conservation constraints or boundary conditions, but in order to identify a SEA path in state space we must  equip the state space with a metric field with respect to which to compute the distance traveled  during the time evolution.

 The generalization is as follows.
  In our previous work, we adopted the proper uniform metric for probability distributions, namely, the Fisher-Rao metric, because in quantum thermodynamics the state representative, the density operator, is essentially a generalized probability distribution. In other frameworks, however, the state representative not always is a probability distribution. Moreover, the present application to the framework of Mesoscopic Non-Equilibrium Thermodynamics \cite{Mazur,BedeauxMazur} shows that standard results such as the Fokker-Planck equation and Onsager theory emerge as straightforward results of SEA dynamics with respect to a metric characterized by a generalized metric tensor that is directly related to the inverse of the generalized conductivity tensor. Since the generalized conductivities represent, at least in the near-equilibrium regime, the strength of the system's reaction when pulled out of equilibrium, it appear that their inverse, i.e., the generalized resistivity tensor, represents the metric with respect to which the time evolution, at least in the near equilibrium, is locally SEA.

  But the local SEA construction does much more, because it offers a strongly thermodynamically consistent way to extend the well-known near-equilibrium theories to the treatment of non-equilibrium states.

  The unified formulation of the local SEA variational problem is as follows and it is not restricted to near equilibrium: \textit{the
  time evolution of the local state is the result of a balance between the effects of transport or Hamiltonian dynamics and the spontaneous and irreversible tendency to advance the local state representative in the direction of maximal  entropy production per unit of distance  traveled in state space compatible with the conservation constraints.}

Geometrically, the measure of distance traveled in state space requires the choice of a local metric tensor.  Physically, the local metric tensor contains the full information about the relaxation kinetics of the material. The standard near-equilibrium results obtain when the local  metric tensor is proportional to the inverse of the local matrix of generalized conductivities, i.e., to the  local generalized resistivity matrix.

  The  structure of the SEA geometrical construction for the description of  highly non-equilibrium dissipative dynamics in the nonlinear domain turns out to be closely related to the  GENERIC \cite{Grmela84,GENERIC1,GENERIC2} formulation of dissipation. The seeds of SEA and GENERIC developed independently in the early 80's with different motivations and approaches, but the common general thrust has been and still is to impose strong thermodynamic consistency  in the dynamical modeling of systems far from stable thermodynamic equilibrium. SEA has focused exclusively on the irreversible, entropy generation component of the time evolution, while GENERIC has emphasized also the  coupling and interplay between the reversible and irreversible components of the time evolution.

  We will show elsewhere \cite{Montefusco14} that the main technical differences are that: (1) SEA chooses a (non-degenerate) Riemannian metric tensor as dissipative structure, while GENERIC chooses two compatible degenerate structures (Poisson and degenerate co-Riemannian); and (2) in the description of a continuum, SEA uses the local entropy density field as potential, while GENERIC   uses the global energy and entropy functionals as potentials. Future work is needed to address also the relationships and establish differences and similarities between the SEA description of far from equilibrium dissipation and other closely related approaches, such as the recent  Contact Geometry of Mesoscopic Thermodynamics and Dynamics \cite{Grmela08,Grmela13,Grmela14}, the general ideas of the  Rate-Controlled Constrained-Equilibrium Approach to Far-From-Local-Equilibrium Thermodynamics \cite{Keck90,Beretta12} and of the Quasi-Equilibrium approximation of Invariant Manifolds \cite{Gorban01}, as well as the works of Ziegler \cite{Ziegler} and Edelen \cite{Edelen}.

The question of what is  \quot{the physical basis} for the SEA scheme (or for that is the same, for the GENERIC scheme) is tricky and in  philosophically ill posed. It is as if one would ask what is the physical basis for believing that a classical system should obey Hamilton's equations or the equivalent minimum action principle. The meaning of \quot{physical reality} is well explained in the classic book on this subject by Henry Margenau \cite{Margenau}. There is a level of perceptions, the empirical world, that we try to describe by defining concepts, their relations  with the plane of perceptions (operational measurement procedures), and relations among concepts  that we call laws or principles (often using the language of mathematics to express them efficiently). The farther the construction goes from the plane of perceptions the more \quot{abstract} it becomes, but the advantage is that more abstraction may allow to encompass and regularize a broader set of less abstract theories, in short, to unify them. At any level of abstraction, what makes a theory \quot{physical} are its links to the plane of perception, namely the fact that the theory allows to model some empirical evidence with some reasonable level of approximation.

Paraphrasing words of Feynman, what makes a particular law or principle \quot{great}, such as the great conservation principles or the second law of thermodynamics, is the fact that they hold for whatever level of description of whatever empirical reality, provided the model has some basic structure and obeys some reasonable conditions, such as those that grant and give meaning to the concept of separability between the object of study and its surroundings.
The spirit of the SEA construction is precisely this. We consider a number of  frameworks that have successfully modeled  non-equilibrium systems at some level of description,  we focus on how these successful models of physical reality describe entropy production by irreversibility, and we cast them in a way that allows us to see that they can all be encompassed and regularized by the unifying geometrical SEA construction.  The GENERIC construction is even more ambitious in that it attempts to unify at once also the reversible and transport contributions by recognizing their common Hamiltonian structure and their relations with the irreversible aspects of the dynamics.

Being more abstract (i.e., farther from Margenau's plane of perceptions) than the various physical theories they unify, the SEA and GENERIC constructions emerge as general dynamical principles which operate within the same domain of validity and hence a similar level of \quot{greatness} of the second law of thermodynamics, by complementing it with the additional essential elements about non-equilibrium behavior.

An important fraction of the greatness of the second law of thermodynamics stems from the fact that it supports the operational definition of entropy \cite{GB2005,ZB2014} as a property of any well-defined system and in any of its equilibrium and non-equilibrium states. Other good fractions that have  direct impact also on the  near-equilibrium description of dynamics derive from the stability and maximal entropy features of the  equilibrium states.

An important fraction of the greatness of the SEA principle stems from the fact that for any  well-defined  system it supports the operational definition  of the metric field $\hat G$ over  its entire state space,  which characterizes even in the far non-equilibrium domain all that can be said about the spontaneous, irreversible, entropy generating tendency towards stable equilibrium. Another good fraction derives from the fact that within the SEA construction the maximum entropy production (MEP) principle acquires a precise and general validity whereby, in any  well-defined model, the entropy producing component of the dynamics effectively pulls the state of the system in the direction of steepest entropy ascent compatible with the metric field $\hat G$ and the imposed conservation laws.

The paper is structured as follows. In Section \ref{sec1} we reformulate
several of the well-known approaches for the description of dissipation in non-equilibrium systems so as to express them all in terms of a common geometrical formulation. In Section \ref{SEA} we then introduce our steepest-entropy-ascent unified variational formulation of non-equilibrium dissipation and discuss its main general features. In Section \ref{sec3} we give a pictorial representation of the same concepts and in Section \ref{concl} we draw our conclusions.

\section{Common  structure of the description of dissipation in several non-equilibrium frameworks}\label{sec1}

In this section we show that
several well-known non-equilibrium frameworks at various  levels of description can be recast in slightly nonstandard, but unifying notation, so that all exhibit as common features the following essential Conditions:
\begin{enumerate}%[leftmargin=*]
\item[C1:] the state space, denoted  by  the symbol  $ \mathcal{L} $, is a manifold in a Hilbert space equipped with a suitable inner product $ ( \cdot | \cdot ) $;
we denote its elements (the states) by $\gamma$ or, alternatively, $|\gamma)$;
\item[C2:] the system properties (energy, entropy, mass, momentum, etc.) are represented by real functionals $\tilde A(\gamma)$, $\tilde B(\gamma)$, \dots  of $\gamma$ such that their functional derivatives with respect to $\gamma$ are also elements of $ \mathcal{L} $; we  denote them by $\delta \tilde A(\gamma)/\delta\gamma$ or, alternatively, by $|\delta \tilde A(\gamma)/\delta\gamma)$;
\item[C3:] if the states are functions of time $t$ only, $\gamma=\gamma(t)$,  their time evolution $\gamma(t)$ obeys the equation of motion \be |d\gamma/dt) = |\Pi_{\gamma}) \label{aa1} \ee where  $|\Pi_{\gamma})$ is also an element  of $ \mathcal{L} $ such that the rates of change of the entropy $\tilde S(\gamma)$ and of any conserved property $\tilde C_i(\gamma)$, with $i$ labeling a list of conserved properties, are \bea && dS/dt = \Pi_S \quad\mbox{with}\quad \Pi_S=  (\Phi|\Pi_{\gamma}) \ge 0 \label{ss1}\\ && dC_i/dt = \Pi_{C_i} \quad\mbox{with}\quad \Pi_{C_i} = (\Psi_i|\Pi_{\gamma}) = 0 \eea where  $\Pi_S$ and $\Pi_{C_i}$ are the respective production rates, and $|\Phi)= | \delta \tilde S(\gamma)/\delta\gamma)$ and $|\Psi_i)= | \delta \tilde C_i(\gamma)/\delta\gamma)$ are shorthand for denoting the variational derivatives with respect to $\gamma$ of the entropy functional $\tilde S(\gamma)$  and  the conserved functional $\tilde C_i(\gamma)$, respectively;
\item[C3':] if the states are continuum fields $\gamma=\gamma(t,x)$, assume that the time evolution obeys the equation of motion \be |\partial \gamma/\partial t)+\mathcal{R}_\gamma|\gamma)=|\Pi_\gamma) \label{aa2} \ee
where $\mathcal{R}_\gamma$ is an operator on $ \mathcal{L} $ responsible for the description of the local fluxes in the continuum and $|\Pi_{\gamma})$ is  an element  of $ \mathcal{L} $  responsible for the description of the local production densities, such that the balance equations for entropy and any conserved property are
\bea && \partial S/\partial t +\nabla\cdot\textbf{J}_S  = \Pi_S \quad\mbox{with}\quad \Pi_S= (\Phi|\Pi_{\gamma}) \ge 0 \label{ss2} \\ && \partial C_i/\partial t +\nabla\cdot\textbf{J}_{C_i}  = \Pi_{C_i} \quad\mbox{with}\quad \Pi_{C_i} = (\Psi|\Pi_{\gamma}) = 0 \label{cc2} \eea
where of course $\textbf{J}_S$ and $\textbf{J}_{C_i}$ are the respective local fluxes, and $\Pi_S$ and $\Pi_{C_i}$ the respective local production densities.
\end{enumerate}

In the next  subsections we introduce the details of the slightly nonstandard notations that allow us to reformulate in the terms just outlined some of the approaches that have been developed over the last several decades to provide thermodynamically consistent theories of non-equilibrium dissipation at various levels of description. This list of approaches is by no means exhaustive and their reformulations have no important elements of novelty. Their presentation is only intended to explicitly substantiate  the above common features in some of the most  well-known non-equilibrium modeling frameworks. Perhaps the only major point is that in order to satisfy Condition C2 in Frameworks A, B, D, and E, we will borrow from the formalism we originally developed
for the quantum framework \cite{Frontiers,LectureNotes} (later introduced also in \cite{Gheorghiu,Reznik}) the use of square-roots of probabilities (instead of the probabilities themselves) as state representative.

 The reader who does not need to be convinced about such details can skip the rest of this section and jump to  Section \ref{SEA} where we provide the unified construction and implementation of  the SEA concept, based only on the general assumptions itemized above.

\subsection{Framework A: Statistical or Information Theoretic Models of Relaxation to Equilibrium}\label{frameA}

Let $ \mathcal{L} $ be the set of all $ n $-vectors of real finite numbers $ A = \vect ( a_j )$, $B = \vect ( b_j )$, \dots
( $n \leq \infty$ ), equipped with the inner product $ ( \cdot |
\cdot ) $ defined by \be \lt A | B \rt = \trac (A B) = {\textstyle
\sum_{j = 1}^n} a_j\, b_j \label{1c} \ee \noindent

In Information Theory \cite{Jaynes}, the probability assignment to a set of $ n $ events, $
p_j $ being the probability of occurrence of the $ j $-th event can be represented by $ \rho = \vect ( p_j ) $. In order to easily impose the constraint of preservation of  nonnegativity of the probabilities during their time evolution  and to obtain Condition C2 above,  we adopt the description in terms of the square-root of $\rho$ that we denote by
\be \gamma = \vect ( \gamma_j=\sqrt{p_j} ) \label{2c}\ee

Typically we consider a set of conserved expectation values of the process \be\lgr \tilde  C_i(\gamma) \rgr= \lgr \tilde  H(\gamma) , \tilde  N_1(\gamma) ,
\ldots , \tilde  N_r(\gamma), \tilde  I(\gamma) \rgr    \ee
where $\tilde  H(\gamma)=\trac (\gamma^2  H)={\textstyle \sum_{j = 1}^n} \gamma_j^2\,e_j $ with $H$ denoting the constant vector $ H = \vect ( e_j )$, for $i=1,\dots,r$,  $\tilde   N_i(\gamma)=\trac (\gamma^2 N_i)={\textstyle \sum_{j = 1}^n} \gamma_j^2\,n_{i j}$ with  $ N_i$ denoting the constant vector $N_i = \vect ( n_{i j } )$,  and $\tilde  I(\gamma)=  \trac (\gamma^2 I)={\textstyle \sum_{j = 1}^n} \gamma_j^2 = 1$, providing the normalization condition, with $I = \vect ( 1 ) $. Notice that the variational derivatives $\delta\tilde   H/\delta \gamma= \vect (2\gamma_j e_j )$, $\delta N_i/\delta \gamma= \vect (2\gamma_j n_{i j} ) $, $\delta I/\delta \gamma= 2\gamma$  are vectors in $
\mathcal{L} $, thus satisfying Condition C2 above.  We denote them collectively by \be \Psi_i=\delta \tilde   C_i/\delta \gamma \ee

A time evolution of the square-root probability distribution,  $\gamma(t)$, is a solution of the rate equation
\be \frac{d\gamma}{dt} = \Pi_{\gamma} \label{3c} \ee
 where the  term $\Pi_{\gamma}$ must be such as to satisfy the constraints of conservation of the expectation values $\tilde  C_i(\gamma ) $, i.e., such that
\be \Pi_{C_i}=\frac{d\tilde   C_i}{dt}=\frac{d}{dt}\trac(\gamma^2 C_i)=   (\Psi_i |\Pi_{\gamma})  = 0 \label{4c} \ee

 The entropy   in this context is represented by  the Shannon functional
\be \tilde  S ( \gamma ) = - k \trac (\rho \ln \rho) =  (-k\gamma\ln \gamma^2 |\gamma) \label{5c} \ee
so that the rate of entropy production is given by
\be \Pi_{S}=\frac{d\tilde  S}{dt}=- k \frac{d}{dt}\trac(\rho \ln \rho)=  (\Phi |\Pi_{\gamma})  \label{6c} \ee
where $\Phi$ denotes its  variational derivative with respect to $\gamma$, \be \Phi=\delta \tilde  S/\delta\gamma=\vect(-2k\gamma_j-2k\gamma_j\ln \gamma_j^2)\ee It is noteworthy that an advantage of the state representation in terms of  square-root probability distributions is that $\delta S/\delta\gamma$ is well defined and belongs to $ \mathcal{L} $ for any distribution, even if some of the probabilities $p_i$ are equal to zero, whereas is such cases $\delta S/\delta\rho$ is undefined and does not belong to $ \mathcal{L} $.

In  Section \ref{SEA} we present the SEA  construction which in this framework  provides a model for the rate term $\Pi_{\gamma}$ whereby  $\Pi_S$ is maximal subject to the conservation constraints $\Pi_{C_i}=0$ and the suitable additional constraint we  discuss therein.

An attempt along the same lines has been presented in \cite{Lemanska}.

\subsection{Framework B: Small-Scale and Rarefied Gases Dynamics}\label{frameB}

Let $ \Omega_c $ be the classical one-particle velocity space, and $ \mathcal{L} $ the set of
real, square-integrable functions $ A , B , \ldots $ on $ \Omega_c
$, equipped with the inner product $ ( \cdot | \cdot ) $ defined
by \be (A | B) = \trac_c (A B) = {\textstyle \int_{\Omega_c}} A B \,\,
d \Omega_c \label{1b} \ee \noindent where $ \trac_c (\cdot) $ in this
framework denotes $ \int_{\Omega_c} \cdot\,d \Omega_c $, with $d \Omega_c =dc_x\,dc_y\,dc_z$.
%
%We denote by
%$ \mathcal{P} $ the subset of all nonnegative-definite distribution
%functions (local one-particle velocity distributions) $ \rho $ in $ \mathcal{L} $, i.e., \be
%\mathcal{P} = \big\{ \rho \mbox{ in } \mathcal{L}\, |\, \rho \geq
%0  \big\} \label{2b} \ee \noindent

In the Kinetic Theory of Rarefied Gases and Small-Scale Hydrodynamics \cite{Nicolas}, the probability to find a particle (at position $\textbf{x}$ and time $t$) with velocity between $\textbf{c}$ and $\textbf{c}+d \textbf{c}$ [where of course $\textbf{c}=(c_x,c_y,c_z)$] is given by  $ f(\textbf{c};\textbf{x},t)\, d \Omega_c / {\textstyle \int_{\Omega_c} } f \, d \Omega_c$ where $ f(\textbf{c};\textbf{x},t) $  is the local phase-density distribution which for every position $\textbf{x}$ and time instant $t$ is a function in  $ \mathcal{L} $.

Also in this framework, in order to easily impose the constraint of preservation of the nonnegativity of $f$ during its time evolution  and to obtain Condition C2,  we introduce the local one-particle state representation not by $f$ itself but by its square root, that we assume is also a function in $ \mathcal{L} $ that we denote by $ \gamma=\gamma(\textbf{c};\textbf{x},t) $.
Therefore, we  have
\be f = \gamma^2 \ ,\quad \frac{\partial f}{\partial t}=2\gamma\frac{\partial\gamma}{\partial t}, \quad \frac{\partial f}{\partial \textbf{x}}=2\gamma\frac{\partial\gamma}{\partial \textbf{x}}\ ,\quad \frac{\partial f}{\partial \textbf{c}}=2\gamma\frac{\partial\gamma}{\partial \textbf{c}} \ee
and for any functionals $\tilde  A_f(f)$ and  $\tilde  A_\gamma(\gamma)=\tilde  A_f(\gamma^2)$
\bea  &\displaystyle\frac{\delta \tilde  A_\gamma( \gamma)}{\delta\gamma }=2\gamma\left.\frac{\delta \tilde  A_f( f)}{\delta f }\right|_{f=\gamma^2}\eea

Again, among the  functionals that represent the one-particle physical observables we focus on the conserved fields, i.e., the collision invariants (mass, momentum, energy), that we denote synthetically by  the set \be\lgr \tilde  C_i(\gamma) \rgr =
\lgr m\,\tilde  n(\gamma), \tilde  M_x(\gamma),  \tilde  M_y(\gamma),  \tilde  M_z(\gamma), \tilde  H(\gamma)\rgr \ee
where $m$ is the single-particle mass, $\tilde  n(\gamma)=\trac_c(\gamma^2)$  the particle number density field, $\tilde  M_i(\gamma)=m\trac_c(\gamma^2 c_i)$  the $i$-th component of the  momentum density field, and the total energy density field $\tilde  H(\gamma)=\tilde  T(\gamma)+\tilde  U(\gamma)$ is in general composed of a kinetic energy contribution $\tilde  T(\gamma)=\onehalf m \trac_c(\gamma^2 \textbf{c}\cdot\textbf{c})$ and a potential energy contribution
$\tilde  U(\gamma)$, such that at position $\textbf{x}$ and time $t$ the functional derivative $\delta \tilde  U(\gamma)/\delta \gamma=2\gamma\varphi_\gamma(\textbf{x},t)$ is a function in $ \mathcal{L} $, thus obeying Condition C2 above, where $\varphi_\gamma(\textbf{x},t)$ is the single-particle potential field.  For example, for a uniform externally applied field in the $z$ direction, $-\nabla\varphi_\gamma(\textbf{x})= \textbf{a}=-a\,\nabla z$ with $a$ constant. Again, for the Vlasov-Poisson kinetic theory \cite{Morrison86}, $\tilde  U(\gamma)= \onehalf \trac_c(\gamma^2 \varphi_\gamma)$ where $\varphi_\gamma(\textbf{x},t)$ is a non-local functional of $\gamma$, $\varphi_\gamma(\textbf{x},t)=\int_{\Omega_{x'}}d \Omega_{x'}\int_{\Omega_{c'}}d \Omega_{c'} V(|\textbf{x}-\textbf{x}'|)\,\gamma^2(\textbf{c}';\textbf{x}',t)$ representing a locally-averaged mean-field single-particle potential due to the effects of the neighboring particles via the interparticle potential  $V$ assumed to be a function of particle distance only.

The dissipative time evolution of the distribution function $f$ is given by the Boltzmann equation or some equivalent simplified kinetic model equation, which in terms of the square-root distribution may be written in the form
\be \frac{\partial\gamma}{\partial t} +\textbf{c}\cdot\nabla\gamma -\nabla\varphi_\gamma\cdot\frac{\partial\gamma}{\partial \textbf{c}} = \Pi_{\gamma}  \label{3b} \ee

In order to satisfy the constraints of  mass, momentum, and energy conservation, the collision term $\Pi_{\gamma}$ must be such that
\be \Pi_{C_i}=   (\Psi_i |\Pi_{\gamma})  = 0 \ee
where
\be \Psi_i=\delta \tilde   C_i/\delta \gamma \ee

The  entropy density field  in this context is represented by
\be S ( \textbf{x},t ) = \tilde  S(\gamma)= - k \trac_c (\gamma^2 \ln \gamma^2) =  (-k\gamma\ln\gamma^2 |\gamma) \label{5b} \ee
the rate of entropy production  is
\be \Pi_{S}=(\Phi |\Pi_{\gamma})   \ee
where
\be \Phi=\delta \tilde   S/\delta \gamma \ee
and the entropy balance equation is
\be - k\frac{\partial \trac(f \ln f)}{\partial t}-k\nabla\cdot\trac(f \,\textbf{c}\,\ln f)= \Pi_{S}   \label{6b} \ee
where $\textbf{J}_S=-k\trac(f \,\textbf{c}\,\ln f)$ represents the  entropy flux field.

In  Section \ref{SEA}, we  construct the family of models for the collision term $\Pi_{\gamma}$ such that $\Pi_S$ is maximal subject to the conservation constraints $\Pi_{C_i}=0$ and the suitable additional geometrical  constraint we  discuss therein.

The resulting  family of SEA kinetic  models of the collision integral
in the Boltzmann equation is currently under investigation by comparing it with  standard models such as the well-known BGK model as well as with Monte Carlo simulations of the original Boltzmann
equation for hard spheres \cite{ASME-Nicolas}. In addition to the strong thermodynamic consistency even far from stable equilibrium, Ref. \cite{ASME-Nicolas} gives a proof that in the near-equilibrium limit the SEA models reduces to the corresponding BGK
models.

In a forthcoming paper \cite{Montefusco14}, we work out the explicit relation between SEA and GENERIC and  we provide the explicit SEA form of the full Boltzmann collision operator by using its already available GENERIC form  \cite{Grmela84,Grmela02,Grmela14,Grmela14b}.

\subsection{Framework C: Rational Extended Thermodynamics, Macroscopic Non-Equilibrium Thermodynamics, and Chemical Kinetics}\label{frameC}

Let $ \mathcal{L} $ be the set of all $ n  $-vectors of real numbers $ A = \vect ( a_j )$, $B = \vect ( b_j )$, \dots
( $n \leq \infty$ ), equipped with the inner product $ ( \cdot |
\cdot ) $ defined by \be \lt A | B \rt = \trac (A B) = {\textstyle
\sum_{j = 1}^n} a_j\, b_j \label{1d} \ee \noindent

In Rational Extended Thermodynamics (RET) \cite{Ruggeri}, the local state at position $\textbf{x}$ and time $t$ of the continuum under study is represented by an element $\gamma$ in $ \mathcal{L} $, i.e.,
\be \gamma(\textbf{x},t)= \vect[\alpha(\textbf{x},t)] \ee

 Thus, $ \gamma(\textbf{x},t) $ represents the  set of fields which represent the instantaneous spatial distributions within the continuum of the local densities that define all its other local properties. In particular, for the  conserved properties energy, momentum, and
  mass \cite{chem1}
 it is assumed  that their local densities and their local (Lagrangian) fluxes  are all given by particular functions of $\gamma$ that we denote synthetically by
 \be\lgr \tilde  C_i (\gamma)  \rgr=
\lgr \tilde  E(\gamma) , \tilde  M_x(\gamma) , \tilde   M_y(\gamma) , \tilde  M_z(\gamma), \tilde  m_k(\gamma) \rgr \ee
\be\lgr \textbf{J}_{C_i}(\gamma)  \rgr =
\lgr  \textbf{J}_E(\gamma) , \textbf{J}_{M_x}(\gamma) , \textbf{J}_{M_y}(\gamma) , \textbf{J}_{M_z}(\gamma), \textbf{J}_{m_k}(\gamma) \rgr \ee
so that the energy, momentum, and mass balance equations take the form
\be \frac{\partial C_i}{\partial t} +\nabla\cdot\textbf{J}_{C_i} = \Pi_{C_i}= 0  \label{3d} \ee
 Moreover, also for the local entropy density and the local (Lagrangian) entropy flux it  is assumed  that they are given by particular functions of $\gamma$ that we denote respectively by
\be S(\gamma)\qquad {\rm and}\qquad \textbf{J}_{S}(\gamma)\ee
so that the entropy balance equation takes the form
\be \frac{\partial S}{\partial t} +\nabla\cdot\textbf{J}_S = \Pi_{S}  \label{4d} \ee
where $\Pi_{S}$ is the local production density.

In general the balance equation for each of the underlying field properties is
\be \frac{\partial \alpha_j}{\partial t}+\nabla\cdot\textbf{J}_{\alpha_j}=\Pi_{\alpha_j} \label{6d} \ee
where $\textbf{J}_{\alpha_j}$ and $\Pi_{\alpha_j}$  are the corresponding flux and production density, respectively. Equivalently, this set of balance equations may be written synthetically  as
\be \frac{\partial \gamma}{\partial t}+\nabla\cdot\textbf{J}_\gamma=\Pi_{\gamma} \label{5d} \ee
where $\textbf{J}_\gamma=\vect (  \textbf{J}_{\alpha_j} )$  and $\Pi_\gamma=\vect ( \Pi_{\alpha_j} )$.

It is then further assumed that there exist  functions $\Phi_{\alpha_j}(\gamma)$ (Liu's Lagrange multipliers \cite{Liu}) that we denote here in vector form by
\be\Phi= \vect( \Phi_{\alpha_j} ) \ee
such that the local entropy production density can be written as
\be \Pi_{S} = {\textstyle
\sum_{j = 1}^n} \Phi_{\alpha_j} \Pi_{\alpha_j}=(\Phi | \Pi_{\gamma})\ee
and must be nonnegative everywhere.

For our development in this paper we  additionally assume that there also exist  functions $\Psi_{i\,{\alpha_j}}(\gamma)$  that we denote  in vector form by
\be\Psi_i= \vect ( \Psi_{i\,{\alpha_j}})  \ee
such that the production density of each conserved property $C_i$ can be written as
\be \Pi_{C_i} = {\textstyle
\sum_{j = 1}^n} \Psi_{i\,{\alpha_j}} \Pi_{\alpha_j}=(\Psi_i | \Pi_{\gamma})\ee

Typically, but not necessarily, the first $4+n_{\rm sp}-n_{\rm r}$ underlying fields $\alpha_j(\textbf{x},t)$ for $j=1,\dots,4+n_{\rm sp}-n_{\rm r}$ are conveniently chosen to coincide with the energy, momentum, and (independently conserved \cite{chem1} linear combinations of the)
 mass densities, where $n_{\rm sp}$ is the number of species and $n_{\rm r}$ the number of independent reactions,
 so that Eqs.\ (\ref{6d}) for  $j=1,\dots,4+n_{\rm sp}-n_{\rm r}$ coincide with Eqs.\ (\ref{3d}) because $\Pi_{\alpha_j}=0$ for this subset of conserved fields.

The above framework reduces to the traditional Onsager theory of macroscopic Non-Equilibrium Thermodynamics (NET) \cite{Mazur} if the $\alpha_j$'s are  taken to represent the local deviations  of the underlying fields from their equilibrium values. In this context, the usual notation calls  the functions $\Phi_{\alpha_j}$ the \quot{thermodynamic forces} and $\Pi_{\alpha_j}$ the \quot{thermodynamic currents}.

In Section \ref{SEA} we  construct an equation of motion for $\gamma$ such that $\Pi_S$ is maximal subject to the conservation constraints $\Pi_{C_i}=0$ plus a suitable additional constraint.

The same framework reduces to the standard scheme of Chemical Kinetics (CK) if the  $\alpha_j$'s include the local reaction coordinates of the $n_{\rm r}$ steps of the detailed kinetic mechanism, the corresponding $\Pi_{\alpha_j}$'s are the local rates of advancement of the reactions, $\Phi_{\alpha_j}=\partial S/\partial \alpha_j =-\sum_{\ell=1}^{n_{\rm sp}}\nu_\ell^j \mu_\ell/T$  is the entropic affinity  of the $j$-th reaction step (equal to the de Donder affinity divided by the temperature, see, e.g.  \cite{Entropy12,JCP04}, where $\mu_\ell$ is the chemical potential of species $\ell$), and $m_k$ are the local values of the $n_{\rm sp}-n_{\rm r}$ independently conserved linear combinations of the
 masses of the various species (see \cite{chem1} for the precise definition) so that $\Pi_{m_k}=0$ are their local production densities.

In the CK framework, the SEA construction is closely related to the gradient-dynamics formulations of the Guldberg--Waage mass action law as  suggested in \cite{Ziegler83,SS87} and more recently in \cite{Grmela12,Klika13}.

 \subsection{Framework D. Mesoscopic Non-Equilibrium Thermodynamics and  Continuum Mechanics with Fluctuations}\label{frameD}

In this section, I renamed variables as follows: what before was  $\pmb{\alpha} = \diag(\alpha_1,\dots,\alpha_m)$ is now $\pmb{\alpha} = \vect(\alpha_1,\dots,\alpha_m) $; what before was $P(\gamma;\textbf{x},t) $ is now $\gamma^2(\pmb{\alpha};\textbf{x},t)$; and the rest accordingly. This is because the state representative is really the probability distribution ( $P(\gamma;\textbf{x},t) $ of the previous version) which we present also here in the square-root form.

Let $ \mathcal{L}$ be the set of all $ n $-vectors $ A = \vect ( a_j(\pmb{\alpha}) )$, $B = \vect ( b_j(\pmb{\alpha}) )$, \dots whose entries $a_j(\pmb{\alpha})$, $b_j(\pmb{\alpha})$, \dots are real, square-integrable functions of a set of mesoscopic variables denoted synthetically by the vector
\be \pmb{\alpha} = \vect(\alpha_1,\dots,\alpha_m) \label{1e} \ee
whose $m$-dimensional range $\Omega_\alpha$ is usually called the  $\pmb{\alpha}$-space.
 Let $ \mathcal{L}$ be
equipped with the inner product $ ( \cdot | \cdot ) $ defined
by \be (A | B) = \sum_{i=1}^n \trac_\alpha (a_i b_i) = \sum_{i=1}^n  \int_{\Omega_\alpha} a_i(\pmb{\alpha}) b_i(\pmb{\alpha}) \,\,
d \Omega_\alpha \label{2e} \ee \noindent where $ \trac (\cdot) $ in this
framework denotes $ \int_{\Omega_\alpha} \cdot\,d \Omega_\alpha $, with $d \Omega_\alpha =d\alpha_1\cdots d\alpha_m$.

 In Mesoscopic Non-Equilibrium Thermodynamics (MNET)  (see, e.g., \cite{Mazur,Rubi})  the  $\alpha_j$'s are the set of mesoscopic (coarse grained) local extensive properties assumed to  represent the local non-equilibrium state of the portion of continuum under study. The mesoscopic  description of the local state at position $\textbf{x}$ and time $t$  is in terms of a square-root probability density  on the $\pmb{\alpha}$-space $\Omega_\alpha$, that we denote by  \begin{equation*}\gamma(\pmb{\alpha};\textbf{x},t)\end{equation*}  such that $\gamma^2(\pmb{\alpha};\textbf{x},t)\,d \Omega_\alpha$ represents the probability that the values of the underlying fields are between $\pmb{\alpha}$ and $\pmb{\alpha}+d\pmb{\alpha}$.

 It is assumed that the probability density $\gamma$  obeys a continuity equation that we may write as follows
 \be \frac{\partial \gamma}{\partial t}+\textbf{v}\cdot  \nabla \gamma=\Pi_\gamma \quad \mbox{ with }\quad  2\gamma\Pi_\gamma=-\nabla_\alpha\cdot \Pi_\alpha \label{3e} \ee
 where $\textbf{v}=\textbf{v}(\pmb{\alpha})$  is the particle velocity expressed in terms of the underlying fields (usually it is convenient to take the first three $\alpha_j$'s to coincide with the velocity components), $\textbf{v}\cdot  \nabla \gamma=\nabla\cdot\textbf{J}_\gamma$ where $\textbf{J}_\gamma$ is the flux of square-root probability density, and
 \be \Pi_\alpha(\pmb{\alpha};\textbf{x},t) =  \vect ( \Pi_{\alpha_j} ) \ , \quad \nabla_\alpha=\vect \left( \frac{\partial}{\partial\alpha_j} \right) \label{4e}\ee
where the $\Pi_{\alpha_j}$'s are interpreted as probability weighted components of a streaming flux in $\Omega_\alpha$, i.e., a current in the space of mesoscopic coordinates.

 The local densities $C_i(\textbf{x},t)$ of the conserved properties are assumed to  have an associated underlying extensive property which can be expressed in terms of the mesoscopic coordinates as  $c_i(\pmb{\alpha})$ such that
 \be C_i(\textbf{x},t)  = \tilde  C_i(\gamma)= \int_{\Omega_\alpha} c_i(\pmb{\alpha})\,\gamma^2(\pmb{\alpha};\textbf{x},t)\,d \Omega_\alpha \ee
 \be \Psi_i=\delta \tilde  C_i/\delta\gamma= 2\gamma c_i \ee
  They obey the balance equation
 \be \frac{\partial C_i}{\partial t} +\nabla\cdot\textbf{J}_{C_i} = \Pi_{C_i}= 0  \label{5e} \ee
  where  the local flux $\textbf{J}_{C_i}(\textbf{x},t)$ and the local production density $\Pi_{C_i}(\textbf{x},t)$ are defined as follows
  \bea
  \textbf{J}_{C_i}(\textbf{x},t)  &=& \int_{\Omega_\alpha} c_i(\pmb{\alpha})\,\textbf{v}(\pmb{\alpha})\,\gamma^2(\pmb{\alpha};\textbf{x},t)\,d \Omega_\alpha \nonumber\\
  \Pi_{C_i}(\textbf{x},t) &=& (\Psi_i| \Pi_\gamma)\nonumber\\ &=& \int_{\Omega_\alpha} c_i(\pmb{\alpha})\,2\gamma (\pmb{\alpha};\textbf{x},t) \Pi_\gamma(\pmb{\alpha};\textbf{x},t)\,d \Omega_\alpha \nonumber\\
 &=&-
  \int_{\Omega_\alpha} c_i(\pmb{\alpha})\,\nabla_\alpha\cdot \Pi_\alpha(\pmb{\alpha};\textbf{x},t)\,d \Omega_\alpha
 \nonumber\\
 &=&
 \int_{\Omega_\alpha} \Pi_\alpha(\pmb{\alpha};\textbf{x},t)\cdot \nabla_\alpha c_i(\pmb{\alpha})\, \,d \Omega_\alpha
 \nonumber\\
 &=& (\psi_i| \Pi_\alpha)
 \label{6e}
  \eea
   where in the next to the last equation we integrated by parts and assumed that currents in $\pmb{\alpha}$-space decay sufficiently fast to zero as the  $\alpha_j$'s $\rightarrow\infty$, and  we defined
  \be \psi_i(\pmb{\alpha})= \nabla_\alpha c_i(\pmb{\alpha}) \label{6e1}\ee

   Also the condition of preservation of normalization is written in the same way, by setting $c_0(\pmb{\alpha})=1$ so that $\Psi_0=2\gamma$ and  the corresponding balance equation (\ref{5e}) with the condition $\Pi_{C_0}(\textbf{x},t)=0$ yields the following conditions on $\Pi_\gamma$ and $\Pi_\alpha$
  \bea
  \Pi_{C_0}(\textbf{x},t) &=& (\Psi_0| \Pi_\gamma)= \int_{\Omega_\alpha} 2\gamma (\pmb{\alpha};\textbf{x},t) \Pi_\gamma(\pmb{\alpha};\textbf{x},t)\,d \Omega_\alpha\nonumber\\ &=&-
  \int_{\Omega_\alpha} \nabla_\alpha\cdot \Pi_\alpha(\pmb{\alpha};\textbf{x},t)\,d \Omega_\alpha =0 \label{norm}\eea

The  local entropy density $S(\textbf{x},t)$ is   expressed in terms of the local square-root probability density as
\be S(\textbf{x},t)  =\tilde  S(\gamma)=-k \int_{\Omega_\gamma} \gamma^2(\pmb{\alpha};\textbf{x},t)\ln \gamma^2(\pmb{\alpha};\textbf{x},t)\,d \Omega_\alpha \ee
such that
\be \Phi=\delta \tilde  S/\delta \gamma= - 2k \gamma(\pmb{\alpha};\textbf{x},t)\,[1+ \ln \gamma^2(\pmb{\alpha};\textbf{x},t)] \ee
 and the
 entropy balance equation takes the form
\be \frac{\partial S}{\partial t} +\nabla\cdot\textbf{J}_S = \Pi_{S}  \label{7e} \ee
where  the local flux $\textbf{J}_{S}(\textbf{x},t)$ and the local production density $\Pi_{S}(\textbf{x},t)$ are defined
 as follows
\bea
  \textbf{J}_{S}(\textbf{x},t)  &=& -k\int_{\Omega_\alpha} \gamma^2(\pmb{\alpha};\textbf{x},t)\,\textbf{v}(\pmb{\alpha})\,\ln \gamma^2(\pmb{\alpha};\textbf{x},t)\,d \Omega_\alpha \nonumber\\
  \Pi_{S}(\textbf{x},t)
&=& (\Phi| \Pi_\gamma)\nonumber\\
 &=& -k\int_{\Omega_\alpha} [1+ \ln \gamma^2(\pmb{\alpha};\textbf{x},t)]\,2\gamma (\pmb{\alpha};\textbf{x},t) \Pi_\gamma(\pmb{\alpha};\textbf{x},t)\,d \Omega_\alpha \nonumber\\
 &=& k  \int_{\Omega_\alpha} \ln \gamma^2(\pmb{\alpha};\textbf{x},t)\,\nabla_\alpha\cdot \Pi_\alpha(\pmb{\alpha};\textbf{x},t)\,d \Omega_\alpha
 \nonumber\\
 &=& -k \int_{\Omega_\alpha} \Pi_\alpha(\pmb{\alpha};\textbf{x},t)\cdot \nabla_\alpha\ln \gamma^2(\pmb{\alpha};\textbf{x},t)\, \,d \Omega_\alpha
\nonumber\\
 &=& (\phi| \Pi_\alpha)
  \label{9e}
  \eea
 where we used the normalization condition (\ref{norm}) and again in the next to the last equation we integrated by parts and  defined
 \be \phi(\pmb{\alpha};\textbf{x},t)= -k\nabla_\alpha\ln \gamma^2(\pmb{\alpha};\textbf{x},t) \label{10e}\ee

  In  Section \ref{SEA}, we  construct an equation of motion for $\gamma$ such that $\Pi_S$ is maximal subject to the conservation constraints $\Pi_{C_i}=0$ and the suitable geometrical constraint we  discuss therein. The result, when introduced in Eq.\ (\ref{3e}) will yield the Fokker-Planck equation for $\gamma(\pmb{\alpha};\textbf{x},t)$ which is also related (see, e.g., \cite{GorbanKarlin}) to the GENERIC structure \cite{Grmela84,GENERIC1,GENERIC2}. The formalism can also be readily extended to the family of Tsallis \cite{Tsallis} entropies in the frameworks of non-extensive thermodynamic models \cite{Gorban}.

\subsection{Framework E: Quantum Statistical Mechanics, Quantum Information Theory, Quantum Thermodynamics, Mesoscopic Non-Equilibrium Quantum Thermodynamics, and Intrinsic Quantum Thermodynamics}\label{frameE}

Let $ \mathcal{H} $ be the Hilbert space (dim $ \mathcal{H} \leq
\infty $) associated with the physical system, and $ \mathcal{L} $ the set of all linear operators $ A
$, $B $, \dots  on $ \mathcal{H} $, equipped with the real inner
product $( \cdot | \cdot )$ defined by \be \lt A | B \rt = \trac
\lt A^{\dag} B + B^{\dag} A \rt/2 \label{1f} \ee \noindent where $
A^{\dag} $ denotes the adjoint of operator $ A $ and $ \trac(
\cdot ) $ the trace functional.

In the quantum frameworks that we consider in this section, the state representative is the density operator $\rho$, i.e.,  a unit-trace, self-adjoint, and nonnegative-definite element of $ \mathcal{L} $.

Instead, also here we will adopt  the state representation in terms of the generalized square root of the density operator, that we developed in this context \cite{Frontiers,LectureNotes,Gheorghiu,ROMP} in order to easily impose the constraints of preservation of both the nonnegativity and the self-adjointness of $\rho$ during its time evolution. Therefore, we assume that the state representative is an element $\gamma$ in $ \mathcal{L} $ from which we can compute the density operator as follows
\be \rho = \gamma\gamma^\dagger \ee
In other words,
 we adopt as state representative not the density operator $\rho$ itself but its generalized square root $\gamma$. Therefore, we clearly have
\be \frac{d\rho}{d t}=\gamma\frac{d\gamma\dagger}{d t}+\frac{d\gamma}{d t}\gamma^\dagger\label{2f}\ee

We  then
consider the set of operators corresponding to the conserved properties, denoted synthetically as \be\lgr C_i \rgr= \lgr H , M_x, M_y, M_z, N_1 , \ldots , N_r, I \rgr \label{3f}\ee Here we assume that these are self-adjoint operators in $ \mathcal{L} $, that each $M_j$ and $ N_i $
commutes with $ H $, i.e., $ H M_j = M_j H $ for  $ j
= x , y , z $ and  $ H N_i = N_i H $ for $ i
= 1 , \ldots , r $, and that $I$ is the identity operator \cite{Note1}.

The semi-empirical description of an irreversible relaxation process is done in this framework by assuming an evolution equation for the state $\gamma$ given by the equations
\bea  \frac{d\gamma}{d t} +\frac{i}{\hbar}\, H\gamma&=& \Pi_\gamma  \label{4f1}\\ \frac{d\gamma^\dagger}{d t} -\frac{i}{\hbar}\,\gamma^\dagger H&=& \Pi_{\gamma^\dagger}  \label{4f2}  \eea
As a result, it is easy to verify that for the density operator the dynamical equation is
\be  \frac{d\rho}{d t} +\frac{i}{\hbar}\, [H,\rho]= \Pi_\gamma\,\gamma^\dagger+\gamma\,\Pi_{\gamma^\dagger} \label{5f}\ee
where $[\cdot,\cdot]$ denotes the commutator. From this
we  see that in order to preserve hermiticity of $\rho$ the dissipative terms  $\Pi_\gamma$ and $\Pi_{\gamma^\dagger}$ must satisfy the conditions
\be \Pi_{\gamma^\dagger}=\Pi^\dagger_\gamma \quad {\rm and}\quad \Pi_{\gamma}=\Pi^\dagger_{\gamma^\dagger}\label{6f}\ee

In order to satisfy the constraints of conservation of the expectation values $\trac (\rho C_i) $, each $C_i$ must commute with $H$, moreover the  term $\Pi_{\gamma}$ must be such that
\be \Pi_{C_i}=\frac{d}{dt}\trac(\rho C_i)= \trac(C_i \Pi_\gamma\,\gamma^\dagger+\gamma\,\Pi_{\gamma^\dagger}C_i)= (2C_i \gamma|\Pi_{\gamma}) = 0 \label{7f} \ee

The entropy functional  in this context is represented by
\be \tilde S ( \gamma ) = - k \trac (\rho \ln \rho) = (-k(\ln \gamma\gamma^\dagger)\, \gamma |\gamma)  \label{8f} \ee
so that the rate of entropy production under a time evolution that preserves the normalization of $\rho$ is given by
\be \Pi_{S}=- k \frac{d}{dt}\trac(\rho \ln \rho)= (-2k(\ln \gamma\gamma^\dagger)\, \gamma |\Pi_{\gamma})  \label{9f} \ee

In Quantum Statistical Mechanics (QSM) and Quantum Information Theory (QIT), $ \rho $ is the von Neumann
statistical or density operator which represents the index of
statistics from a generally heterogeneous ensemble of identical
systems (same Hilbert space $ \mathcal{H} $ and operators $\lgr H
, N_1 , \ldots , N_r \rgr$) distributed over a range of generally
different quantum mechanical  states. If each individual member of
the ensemble is isolated and uncorrelated from the rest of the
universe, its state is described according to Quantum Mechanics by
an idempotent density operator
($\rho^2=\rho=P_{|\psi\rangle}=\frac{|\psi\rangle\langle\psi|}{\langle\psi|\psi\rangle}$),
i.e., a projection operator onto the span of some vector
$|\psi\rangle$ in $ \mathcal{H} $. If the ensemble is
heterogeneous, its individual member systems may be in different
states, $P_{|\psi_1\rangle}$, $P_{|\psi_2\rangle}$, and so on, and the ensemble statistics is captured by the von Neumann statistical operator $\rho=\sum_j w_j P_{|\psi_j\rangle}$. The entropy functional here represents a measure of the informational uncertainty as to  which homogeneous subensemble the next system will be drawn from, i.e., as to which will be the actual pure quantum state among those present in the heterogeneous ensemble.

In this framework, unless the statistical weights $w_j$ change for some extrinsic reason, the quantum evolution of the ensemble is given by Eq.\ (\ref{5f}) with $\Pi_\gamma=0$ so that
 Eq.\ (\ref{5f}) reduces to von Neumann's equation of quantum (reversible) Hamiltonian evolution, corresponding to $\rho(t)=\sum_j w_j P_{|\psi_j(t)\rangle}$ where the underlying pure states $ |\psi_j(t)\rangle$ evolve according to the Schr\"{o}dinger equation $d |\psi_j\rangle/dt=-iH|\psi_j\rangle/\hbar$.

In the framework of QSM and QIT, the SEA  equation of motion we  construct in the next Section \ref{SEA} for $\rho$ represents a model for the rates of change of the statistical weights $w_j$ in  such a way that $\Pi_S$ is maximal subject to the conservation constraints $\Pi_{C_i}=0$ (and a suitable additional constraint, see Section \ref{SEA}). This essentially extends to the quantum landscape the same statistical or information theoretic non-equilibrium problem we defined above as Framework A.

In Quantum Thermodynamics (QT), instead, the density operator takes on a more fundamental physical meaning. It is not any longer related to the heterogeneity of the ensemble, and it is not any longer assumed that the individual member systems of the ensemble are in pure states.

 The prevailing interpretation of QT  (for a recent review see \cite{vonSpakovskyGemmer}) is the so-called open-system model whereby the quantum system under study (each individual system of a homogeneous ensemble) is always viewed as in contact (weak or strong) with a thermal reservoir or 'heat bath', and its not being in a pure state is an indication of its being correlated with the reservoir. The overall system-plus-reservoir composite is assumed to be in a pure quantum mechanical state $ \mathcal{H}\otimes \mathcal{H}_R $ and reduces to  the density operator $\rho$ on the system's space $ \mathcal{H}$ when we partial trace the overall density operator over the reservoir's space $ \mathcal{H}_R$.

The semi-empirical description of an irreversible relaxation process is done in this framework by assuming for $\Pi_\rho$ in Eq.\ (\ref{5f}) the  Lindblad-Gorini-Kossakowski-Sudarshan (LGKS) form \cite{Lindblad,Kossakowski}
 \be \Pi_\rho =\sum_j \left(V_j\rho V^\dagger_j-\onehalf\{V^\dagger_j V_j,\rho \}_+\right) \label{10f} \ee
 where  $\{\cdot,\cdot\}_+$  denotes the anticommutator and operators $V_j$ are to be chosen so as to properly model the system-reservoir interaction. The justification and modeling assumptions that lead to the general form of Eq.\ (\ref{10f}) are well known.

 In the framework of QT the SEA  equation of motion we  construct in the next section for $\rho$ may be useful as an alternative  model for $\Pi_\rho$ (or for a term additional to the LGKS term) such that $\Pi_S$ is maximal subject to the conservation constraints $\Pi_{C_i}=0$ (and the suitable additional constraint defined below in Section \ref{SEA}). In some cases this could be  simpler than the LGKS model and it has the advantage of a strong built-in thermodynamics consistency. A similar attempt has been recently discussed in Ref. \cite{OttingerQuantum} as an application of the GENERIC scheme.

Mesoscopic Non-Equilibrium Quantum Thermodynamics (MNEQT) \cite{BedeauxMazur} starts from the formalism of QSM but attempts to extend the Onsager NET theory and  MNET to the quantum realm. We will show elsewhere that the present SEA formulation reduces to MNEQT in the near-equilibrium limit, and can therefore be viewed as the natural extension of MNEQT to the far-non-equilibrium regime. The essential elements of this proof have actually already been given \cite{Gheorghiu}, but only for  the particular case corresponding to Eq. (\ref{4g}) below (Fisher-Rao metric).

 An even more fundamental physical meaning is assumed within the theory that we originally called Quantum Thermodynamics \cite{Frontiers,LectureNotes,HG,thesis,Cimento1,Cimento2,JPconf} but more recently renamed Intrinsic Quantum Thermodynamics (IQT) to avoid confusion with the more traditional theories of QT such as those just outlined.

IQT assumes that the second law of thermodynamics should complement the laws of mechanics even at the single particle level \cite{HG}. This can be done if we accept that the true individual quantum state of a system, even if fully isolated and uncorrelated from the rest of the
universe, requires density operators $\rho$ that are not
necessarily idempotent. Over the set of idempotent $\rho$'s, QT
coincides with Quantum Mechanics (QM), but it differs fundamentally
from QM because it assumes a broader set of possible states,
corresponding to the set of non-idempotent $\rho$'s. This way, the entropy
functional $\tilde S(\rho)$ becomes in IQT an intrinsic fundamental property. In a sense IQT with its SEA dynamical law accomplishes the conceptual program, so intensely sought for also by Ilya Prigogine and coworkers \cite{Prigogine}, of answering the following questions \cite{Frontiers}: \textit{What if entropy, rather than a statistical, information theoretic, macroscopic or phenomenological concept, were an intrinsic property of matter in the same sense as energy is universally understood to be an intrinsic property of matter? What if irreversibility were an intrinsic feature of the fundamental dynamical laws obeyed by all physical objects, macroscopic and microscopic, complex and simple, large and small? What if the second law of thermodynamics, in the hierarchy of physical laws, were at the same level as the fundamental laws of mechanics, such as the great conservation principles?}  When viewed from such extreme perspective, the IQT conceptual scheme remains today as \quot{adventurous} as it was acutely judged by John Maddox in 1985 \cite{Nature}.

 In the framework of IQT the SEA  equation of motion (\ref{5f}) for $\rho$ which results from the expression for $\Pi_\gamma$ we  construct in the next section represents a strong family of implementations of the  MEP principle at the fundamental quantum level which contains our original formulation as a special case.

 Even the brief discussion above shows clearly that the differences between QSM, QIT, QT, IQT, and MNEQT are important on the interpretational and conceptual levels. Nevertheless, it is also clear that they all share the same basic mathematical framework. Hence, we believe that the SEA dynamical model, which we show here fits their common mathematical basis, can find in the different theories different physical interpretations and applications.

 \section{Steepest-Entropy-Ascent Dynamics. Unified Variational Formulation of Non-Equilibrium Dissipation}\label{SEA}

 In the preceding section we formulated the non-equilibrium problem in various different frameworks in a unifying way that allows us to represent their dissipative parts in a single formal way. In essence,   as summarized by Conditions C1--C4 above,  the state is represented by an element $\gamma$ of a suitable vector space $ \mathcal{L} $ equipped with an inner product $ (\cdot|\cdot)$. The term in the dynamical equation for $\gamma$ which is responsible for  dissipative irreversible relaxation and hence entropy generation is another element $\Pi_\gamma$ of $ \mathcal{L} $ which,  together with the variational derivatives $\Phi$ and $\Psi_i$ of the functionals $\tilde  S(\gamma)$ and $\tilde  C_i(\gamma)$ representing respectively the entropy and the constants of the motion,   determines the rate of entropy production according to the relation
 \be \Pi_S=(\Phi|\Pi_\gamma) \label{1g} \ee
 and the rates of production of the conserved properties $C_i$ according to the relation
  \be \Pi_{C_i}=(\Psi_i|\Pi_\gamma) \label{2g} \ee

 The formulations in terms of square roots of probabilities in Framework A, of the square root of the phase density in Framework B, of the square-root
probability density in Framework D, of the generalized square root of the density operator in Framework F take care  not only  of the important condition that for the evolution law to be well defined it must conserve the nonnegativity of probabilities, phase densities and density operators (which must also remain self adjoint),  but also of Condition C2, namely, that functional derivatives of the entropy and the constants of the motion are also elements of the vector space $ \mathcal{L} $.

  We are now ready to formulate the SEA construction. We do this by assuming that the time evolution of the state $\gamma$ follows the path of steepest entropy ascent in $ \mathcal{L} $ compatible with the constraints. So, for any given state $\gamma$, we must find the $\Pi_\gamma$ which maximizes the entropy production $\Pi_S$ subject to the constraints $\Pi_{C_i}=0$. But in order to identify the SEA path we are not interested in the unconditional increase in  $\Pi_S$ that we can trivially obtain by simply increasing the \quot{norm} of $\Pi_\gamma$ while keeping its direction fixed. Rather, the SEA path is identified by the direction of $\Pi_\gamma$ which maximizes $\Pi_S$ subject to the constraints, regardless of the norm of $\Pi_\gamma$. Hence, we must do the maximization at constant  norm of $\Pi_\gamma$.

  In the absence of Hamiltonian or transport contributions to the time evolution of $\gamma$, the vector $\Pi_\gamma$ is tangent to the path $\gamma(t)$. Therefore,
the norm of $\Pi_\gamma$ represents the square of the distance $d\ell$ traveled by $\gamma$ in the state space $ \mathcal{L} $ in the time interval $dt$, the square of the \quot{length} of the infinitesimal bit of path traveled in state space in the interval $dt$. The variational problem that identifies the SEA direction at each state $\gamma$ looks at all possible paths through $\gamma$, each characterized by a possible choice for $\Pi_\gamma$. Among all these paths it selects the one with the highest  entropy produced in the interval $dt$, $\Pi_S\,dt$ per unit of distance $d\ell$ traveled by $\gamma$.

  It is therefore apparent that we cannot identify a SEA path until we equip the space $ \mathcal{L} $ with a metric field with respect to which to compute the distance $d\ell$ traveled and the norm of $\Pi_\gamma$.

  In our previous work \cite{ROMP},  we selected the Fisher-Rao metric based on the inner product $(\cdot|\cdot)$ defined on $ \mathcal{L} $. Indeed, in dealing with probability distributions it has been argued by several authors that the Fisher-Rao metric is the proper unique metric for the purpose of computing the distance between two probability distributions (see e.g. \cite{Wootters,Salamon,Braunstein}). According to this metric, the distance between two states $\gamma_1$ and $\gamma_2$ is given by \be d(\gamma_1,\gamma_2)=\sqrt{2}\arccos(\gamma_1|\gamma_2) \label{3g}\ee which implies that the distance traveled along a trajectory in state space is \be d\ell = 2\sqrt{(\Pi_\gamma|\Pi_\gamma)}\, dt \label{4g}\ee As a result, for Framework E the SEA dynamics we have originally proposed is most straightforward.

  However, here we will not adopt a priori a specific metric but rather assume a most general metric, which in Framework E generalizes our previous work and in the other frameworks provides the most general formulation. We assume the following expression for the distance traveled along a short bit of trajectory in state space  \be d\ell = \sqrt{(\Pi_\gamma|\,\hat G(\gamma)\,|\Pi_\gamma)}\, dt \label{5g}\ee
  where $\hat G(\gamma)$ is a real, symmetric, and positive-definite operator on $ \mathcal{L} $ that we call the metric tensor field,  (super)matrix, or (super)operator depending on the framework.  In general $\hat G(\gamma)$ may be a nonlinear function of $\gamma$.  In Framework E, since $ \mathcal{L} $ is the space of operators on the Hilbert space $\mathcal{H}$ of the quantum system, $\hat G$ is a superoperator on $\mathcal{H}$. However, a simple case is when $\hat G|A)=|GA)$ with $G$ some self-adjoint positive-definite operator in $ \mathcal{L} $.

   We may now finally state \textit{the SEA variational problem} and solve it. The problem is to \textit{find the instantaneous \quot{direction} of
   $\Pi_\gamma$ which maximizes the entropy production rate $\Pi_S$ subject to the constraints $\Pi_{C_i}=0$.} We solve it by maximizing the entropy production rate $\Pi_S$ subject to the constraints $\Pi_{C_i}=0$ and the additional constraint $(d\ell/dt)^2=\dot\epsilon^2=$ prescribed. The last constraint keeps the norm of $\Pi_\gamma$ constant as necessary in order to  maximize only with respect to its direction.  From Eq.\ (\ref{5g})  it amounts to keeping fixed the value of $(\Pi_\gamma|\,\hat G\,|\Pi_\gamma)$ at some small positive constant $\dot\epsilon^2$. The solution is easily obtained by the method of Lagrange multipliers. We seek the unconstrained maximum, with respect to $\Pi_\gamma$, of the Lagrangian
     \be  \Upsilon= \Pi_S - \sum_i \beta_i\, \Pi_{C_i} - \frac{\tau}{2}\, (\Pi_\gamma|\,\hat G\,|\Pi_\gamma) \label{8g1} \ee
     where $\beta_i$ and $\tau/2$ are the Lagrange multipliers.  Like $ \hat G$,  they must be independent of $\Pi_\gamma$ but can be functions of the state $\gamma$.
Using Eqs.\ (\ref{1g}) and (\ref{2g}), we rewrite (\ref{8g1}) as follows
   \be  \Upsilon= (\Phi|\Pi_\gamma) - \sum_i \beta_i\, (\Psi_i|\Pi_\gamma) - \frac{\tau}{2}\,  (\Pi_\gamma|\,\hat G\,|\Pi_\gamma) \label{8g} \ee
Taking the variational derivative of $\Upsilon$ with respect to $|\Pi_\gamma)$ and setting it equal to zero we obtain
  \be  \frac{\delta\Upsilon}{\delta \Pi_\gamma} = |\Phi) -\sum_i \beta_i\, |\Psi_i) -\tau \hat G|\Pi_\gamma) =0 \label{9g} \ee
  where we used the identity $(\Pi_\gamma|\,\hat G=\hat G\,|\Pi_\gamma) $ which follows from the symmetry of $\hat G$.
  Thus, we obtain the SEA general evolution equation (the main result of this paper)
  \be |\Pi_\gamma)=\hat L\,|\Phi -\sum_j \beta_j\, \Psi_j)\label{10g} \ee
  where we define for convenience
  \be \hat L=\frac{1}{\tau}\hat G^{-1} \label{10g1}\ee
  Since in the various frameworks $\hat L$ can be connected with the generalized Onsager conductivity (super)matrix in the near equilibrium regime, we see here that  $\tau \hat L$ is the inverse of the metric (super)matrix $\hat G$ with respect to which the dynamics is SEA. In other words, denoting the generalized Onsager resistivity (super)matrix by $ \hat R$ we have: $ \hat R$ = $\tau\,\hat G$. Since, $\hat G$ is positive definite and symmetric, so are $\hat L$ and $ \hat R$. In other words, the SEA assumption  automatically  entails Onsager reciprocity  near thermodynamic equilibrium.

  Inserting Eq.\ (\ref{10g}) into the conservation constraints (\ref{2g}) yields the important system of equations which defines the values of the Lagrange multipliers $\beta_j$,
  \be  \sum_j  (\Psi_i|\,\hat L\,|\Psi_j)\,\beta_j =(\Psi_i|\,\hat L\,|\Phi)\label{11g} \ee
This system can be readily solved for the $\beta_j$'s (for example by Cramer's rule)  because the  functionals $(\Psi_i|\hat L|\Psi_j)$ and $(\Psi_i|\hat L|\Phi)$ are readily computable for the current state $ \gamma$. Notice that the determinant of the matrix $[(\Psi_i|\,\hat L\,|\Psi_j) ]$ is a Gram determinant and its being positive definite is equivalent to the condition of linear independence of the conservation constraints.
 When Cramer's rule is worked out explicitly, the SEA equation (\ref{10g}) takes the form of a ratio of determinants with which we  presented it in the IQT framework \cite{ROMP,thesis,Cimento1,Cimento2,JPconf},  namely,
 \begin{equation}\label{SEAgram} |\Pi_\gamma)=
%\resizebox{.8\hsize}{!}{$
\frac{\left|
\begin{array}{cccc} \hat L\Phi & \hat L\Psi_1 & \cdots& \hat L\Psi_n \\ \\
(\Psi_1|\,\hat L\,|\Phi) & (\Psi_1|\,\hat L\,|\Psi_1)  & \cdots & (\Psi_1|\,\hat L\,|\Psi_n) \\ \\
\vdots & \vdots  & \ddots & \vdots \\ \\
(\Psi_n|\,\hat L\,|\Phi) & (\Psi_n|\,\hat L\,|\Psi_1)  & \cdots & (\Psi_n|\,\hat L\,|\Psi_n)
\end{array} \right|}{\left|
\begin{array}{ccc}
(\Psi_1|\,\hat L\,|\Psi_1)  & \cdots & (\Psi_1|\,\hat L\,|\Psi_n) \\ \\
 \vdots  & \ddots & \vdots \\ \\
 (\Psi_n|\,\hat L\,|\Psi_1)  & \cdots & (\Psi_n|\,\hat L\,|\Psi_n)
 \end{array} \right|}
 %$}
\end{equation}
where the set of vectors $\hat L^{1/2}\,|\Psi_1),\dots,\hat L^{1/2}\,|\Psi_n)$ are linearly independent so that the Gram determinant at the denominator is strictly positive. These are all the vectors in the set $\{\hat L^{1/2}\,|\Psi_i)\}$ if they are linearly independent, otherwise they are a subset of $n$ of them which are linearly independent.

 We can now immediately prove the general consistence with the thermodynamic principle of entropy non-decrease ($H$-theorem in Framework B). Indeed, subtracting  Eqs.\ (\ref{2g}) each multiplied by the corresponding $\beta_j$  from Eq.\ (\ref{1g}) and then inserting Eq.\ (\ref{10g}) yields the following explicit expression for the rate of entropy production
  \bea \Pi_S&=&(\Phi|\Pi_\gamma) = (\Phi-\sum_j \beta_j\, \Psi_j |\Pi_\gamma) \nonumber\\
  &=&(\Phi-\sum_i \beta_i\, \Psi_i |\,\hat L\,|\Phi -\sum_j \beta_j\, \Psi_j)\ge 0 \label{12g}\eea
which is clearly nonnegative-definite by virtue, again, of the nonnegativity that must be assumed for a well defined metric superoperator $\hat G$.

It is interesting to write the expression for  the (prescribed) speed $d\ell/dt$ at which the state $\gamma$ evolves along the SEA path.  This amounts to inserting Eq.\ (\ref{10g}) into the additional constraint $(d\ell/dt)^2=\dot\epsilon^2=$ prescribed. We readily find
\bea \frac{d\ell^2}{dt^2}&=&(\Pi_\gamma|\,\hat G\,|\Pi_\gamma) \nonumber\\
  &=&\frac{1}{\tau^2}(\Phi-\sum_i \beta_i\, \Psi_i |\,\hat G^{-1}\hat G\hat G^{-1}\,|\Phi -\sum_j \beta_j\, \Psi_j)\label{13g1}\\ &=&\frac{1}{\tau}\Pi_S = \dot\epsilon^2\label{13g} \eea
 so that we have the relations
  \bea \tau&=&\frac{\sqrt{(\Phi-\sum_i \beta_i\, \Psi_i |\,\hat G^{-1}\,|\Phi -\sum_j \beta_j\, \Psi_j)}}{d\ell/dt}\label{14g1}\\ &=&\frac{(\Phi-\sum_i \beta_i\, \Psi_i |\,\hat G^{-1}\,|\Phi -\sum_j \beta_j\, \Psi_j)}{\Pi_S}\label{14g} \eea
from which we see that through the Lagrange multiplier $\tau$  we may specify either the speed at which $\gamma$ evolves along the SEA trajectory in state space or the instantaneous rate of entropy production.
 Hence, using $\tau$ given by Eq.\ (\ref{14g}) the evolution equation   (\ref{10g}) will produce a SEA trajectory in state space with the prescribed entropy production $\Pi_S$.
 These relations also  support the interpretation of $\tau$ as the \quot{overall relaxation time}. We see this as follows.

In general, we may  interpret the vector
\be |\Lambda)= \hat G^{-1/2}\,|\Phi-\sum_i \beta_i\, \Psi_i)\label{18g}\ee
as a vector of \quot{non-equilibrium affinities} or, more precisely, of \quot{generalized partial affinities}. In terms of this vector, Eq.\ (\ref{10g}) rewrites as
 \be\hat G^{1/2}\, |\Pi_\gamma)=\frac{1}{\tau}\,|\Lambda)\label{19} \ee
When only some of the partial affinities in the vector $\Lambda$ are zero, the state is partially equilibrated (equilibrated with respect to the corresponding underlying components of the state $\gamma$). When the entries of the vector $\Lambda$  are all zero, then and only then we have an equilibrium state or a non-dissipative limit cycle. In fact, that is when and only when the entropy production vanishes. $(\Lambda|\Lambda)$, which with respect to the metric tensor $\hat G$ is the norm of the vector  $|\Phi -\sum_j \beta_j\, \Psi_j)$, represents a measure of the \quot{overall degree of disequilibrium} of the state $\gamma$. It is important to note that this definition is valid no matter how far the state is from the (maximum entropy) stable equilibrium state, i.e., also for highly non-equilibrium states.

We have proved in the IQT framework, and the result can be readily extended to all other frameworks, that among the equilibrium states only the maximum entropy one is not unstable (in the sense of Lyapunov \cite{Lyapunov}). As a
result, the maximum entropy states emerge as the only stable equilibrium
ones and, therefore, we can  assert that  the SEA construction implements the Hatsopoulos-Keenan statement of the
second law \cite{HK65,GB2005}  at the level of description of everyone of the frameworks we are considering.

 Eq.\ (\ref{14g}) rewrites as
\be \Pi_S = \frac{(\Lambda|\Lambda)}{\tau}\label{20g}\ee
which shows that the rate of entropy production is proportional to the overall degree of disequilibrium. The relaxation time $\tau$ may be a state functional and needs not be constant, but even if it is, the SEA principle provides a  nontrivial non-linear evolution equation that is well defined and reasonable, i.e., thermodynamically consistent, even far from equilibrium.

 We finally note that when the only contribution to the entropy change comes from the production term $\Pi_S$ (for example in Framework B in the case of homogeneous relaxation in the absence of entropy fluxes, or in Framework E for an isolated system), i.e., when the entropy balance equation  reduces to $dS/dt=\Pi_S$, Eq.\ (\ref{13g} ) may be rewritten as
  \be \frac{d\ell}{dt/\tau}=\frac{dS}{d\ell}\label{15g} \ee
  from which we see that when time $t$ is measured in units of $\tau$ the "speed" along the SEA trajectory  is equal to the local rate of entropy increase  along the trajectory.

  If the state $\gamma$ moves only due to the dissipative term $\Pi_\gamma$ (for example in Framework E when $[H,\gamma\gamma^\dagger]=0$), then the overall length of the trajectory in state space traveled between $t=0$ and $t$ is  given by
  \be \ell(t) = \int_0^t \sqrt{(\Pi_\gamma|\,\hat G\,|\Pi_\gamma)}\, dt \label{16g}\ee
  and, correspondingly, we may also define the \quot{non-equilibrium action}
   \be \Sigma = \frac{1}{2}\int_0^t (\Pi_\gamma|\,\hat G\,|\Pi_\gamma)\, dt =\frac{1}{2}\int_0^t \frac{\Pi_S}{\tau}\, dt =\frac{1}{2}\int_0^t \frac{(\Lambda|\Lambda)}{\tau^2}\, dt\label{17g}\ee
   where for the last two equalities we used Eq.\ (\ref{13g}) and Eq.\ (\ref{20g}), respectively.

\section{Pictorial Representations}\label{sec3}

Let us  give pictorial representations of the vectors that we defined in the SEA construction. We consider first the simplest scenario of a uniform metric tensor $\hat G = \hat I$.

\begin{figure}[t]
      \centering
       \includegraphics[width=0.5\textwidth]{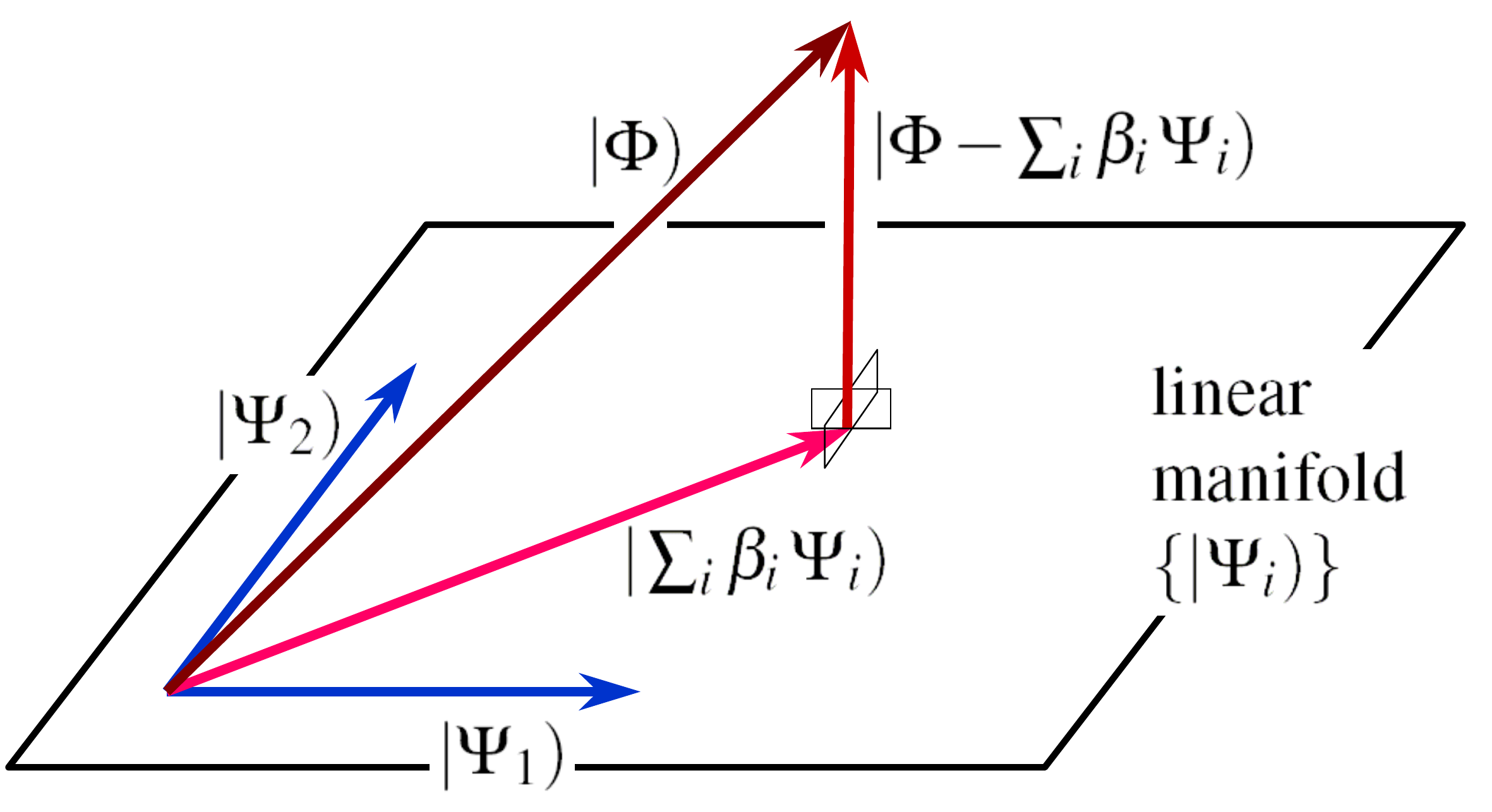}
       \caption{\label{Figure1}Pictorial representation of the linear manifold spanned by the vectors $|\Psi_i)$ and the orthogonal projection of $|\Phi)$ onto this manifold which defines the Lagrange multipliers $\beta_i$ in the case of a uniform metric $\hat G = \hat I$. The construction defines also the generalized affinity vector, which in this case is $|\Lambda)=| \Phi-\sum_i \beta_i\, \Psi_i)$.}
   \end{figure}

 \begin{figure}[t]
      \centering
       \includegraphics[width=0.5\textwidth]{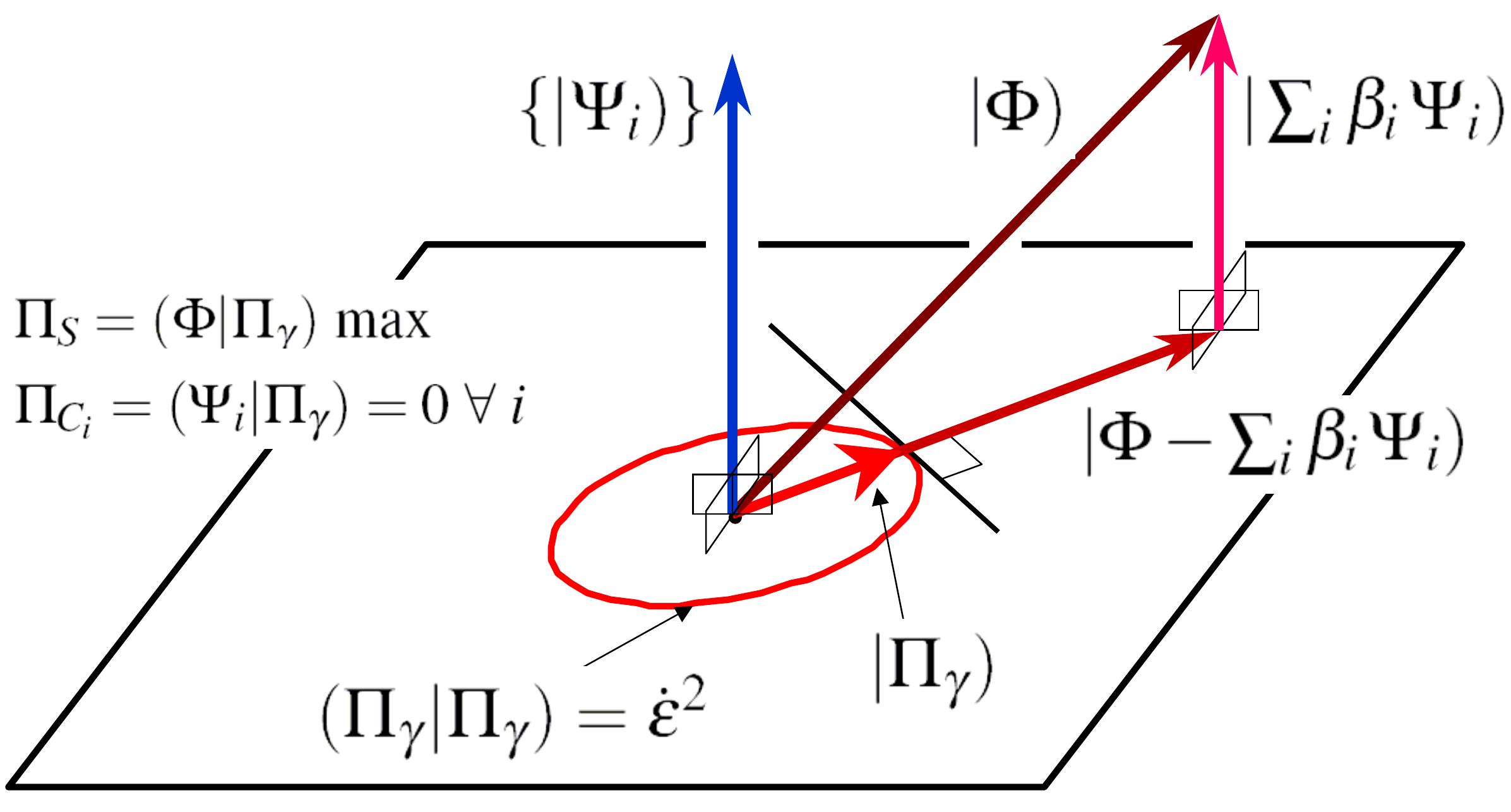}
       \caption{\label{Figure2}Pictorial representation of the SEA variational construction in the case of a uniform metric $\hat G = \hat I$. The circle represents the condition $(\Pi_\gamma|\Pi_\gamma)=\dot\epsilon^2$. The vector $|\Pi_\gamma)$ must be orthogonal to the $|\Psi_i)$'s in order to satisfy the conservation constraints $\Pi_{C_i}=(\Psi_i|\Pi_\gamma)=0$. In order to  maximize the scalar product  $(\Phi-\sum_i \beta_i\, \Psi_i|\Pi_\gamma)$, $|\Pi_\gamma)$ must have the same direction as $| \Phi-\sum_i \beta_i\, \Psi_i)$.}
   \end{figure}

Figure \ref{Figure1} gives a pictorial representation of the linear manifold spanned by the vectors $|\Psi_i)$'s and the orthogonal projection of $|\Phi)$ which defines the Lagrange multipliers $\beta_i$ in the case of uniform metric, i.e., the orthogonality conditions  $(\Psi_j|\Phi-\sum_i \beta_i\, \Psi_i)=0$ for every $j$, which is Eq.\ (\ref{11g}) with $\hat L = \hat I/\tau$. The construction defines also the generalized affinity vector, which in this case is $|\Lambda)=| \Phi-\sum_i \beta_i\, \Psi_i)$ and is orthogonal to the linear manifold spanned by the vectors $|\Psi_i)$'s.

Figure \ref{Figure2} gives a pictorial representation of the subspace orthogonal to the linear manifold spanned by the $|\Psi_i)$'s that here we denote for simplicity by $\{ |\Psi_i) \}$. The vector $|\Phi)$ is decomposed into its component $|\sum_i \beta_i\, \Psi_i)$ which lies in $\{ |\Psi_i) \}$  and its component $|\Phi-\sum_i \beta_i\, \Psi_i)$ which lies in the orthogonal  subspace.

The circle in Figure \ref{Figure2} represents the condition $(\Pi_\gamma|\Pi_\gamma)=\dot\epsilon^2$ corresponding in the uniform metric to the  prescribed rate of advancement in state space, $\dot\epsilon^2=(d\ell/dt)^2$. The compatibility with the conservation constraints $\Pi_{C_i}=(\Psi_i|\Pi_\gamma)=0$ requires that  $|\Pi_\gamma)$ lies in the subspace orthogonal to the  $|\Psi_i)$'s. To take the SEA
 direction, $|\Pi_\gamma)$ must maximize the scalar product  $(\Phi-\sum_i \beta_i\, \Psi_i|\Pi_\gamma)$. This clearly happens when $|\Pi_\gamma)$ has the same direction as the vector $| \Phi-\sum_i \beta_i\, \Psi_i)$ which in the uniform metric coincides with  the generalized affinity vector $|\Lambda)$.

   \begin{figure}[t]
      \centering
       \includegraphics[width=0.5\textwidth]{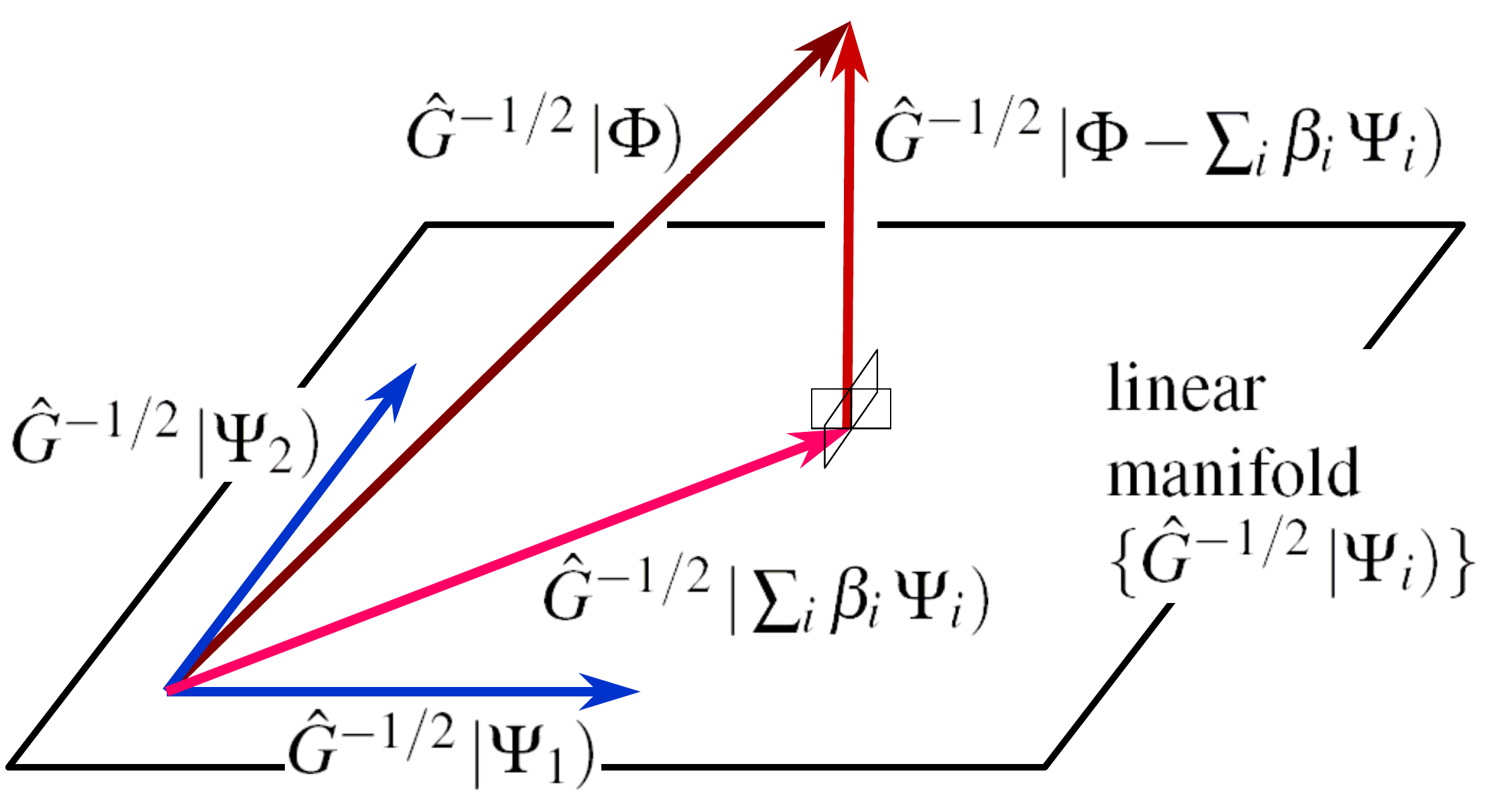}
       \caption{\label{Figure3}Pictorial representation of the linear manifold spanned by the vectors $\hat G^{-1/2}\,\Psi_i$ and the orthogonal projection of $\hat G^{-1/2}\,|\Phi)$ onto this manifold which defines the Lagrange multipliers $\beta_i$ in the case of  a non-uniform metric $\hat G$. The construction defines also the generalized affinity vector $|\Lambda)=\hat G^{-1/2}\,| \Phi-\sum_i \beta_i\, \Psi_i)$.}
   \end{figure}
 \begin{figure}[t]
      \centering
       \includegraphics[width=0.5\textwidth]{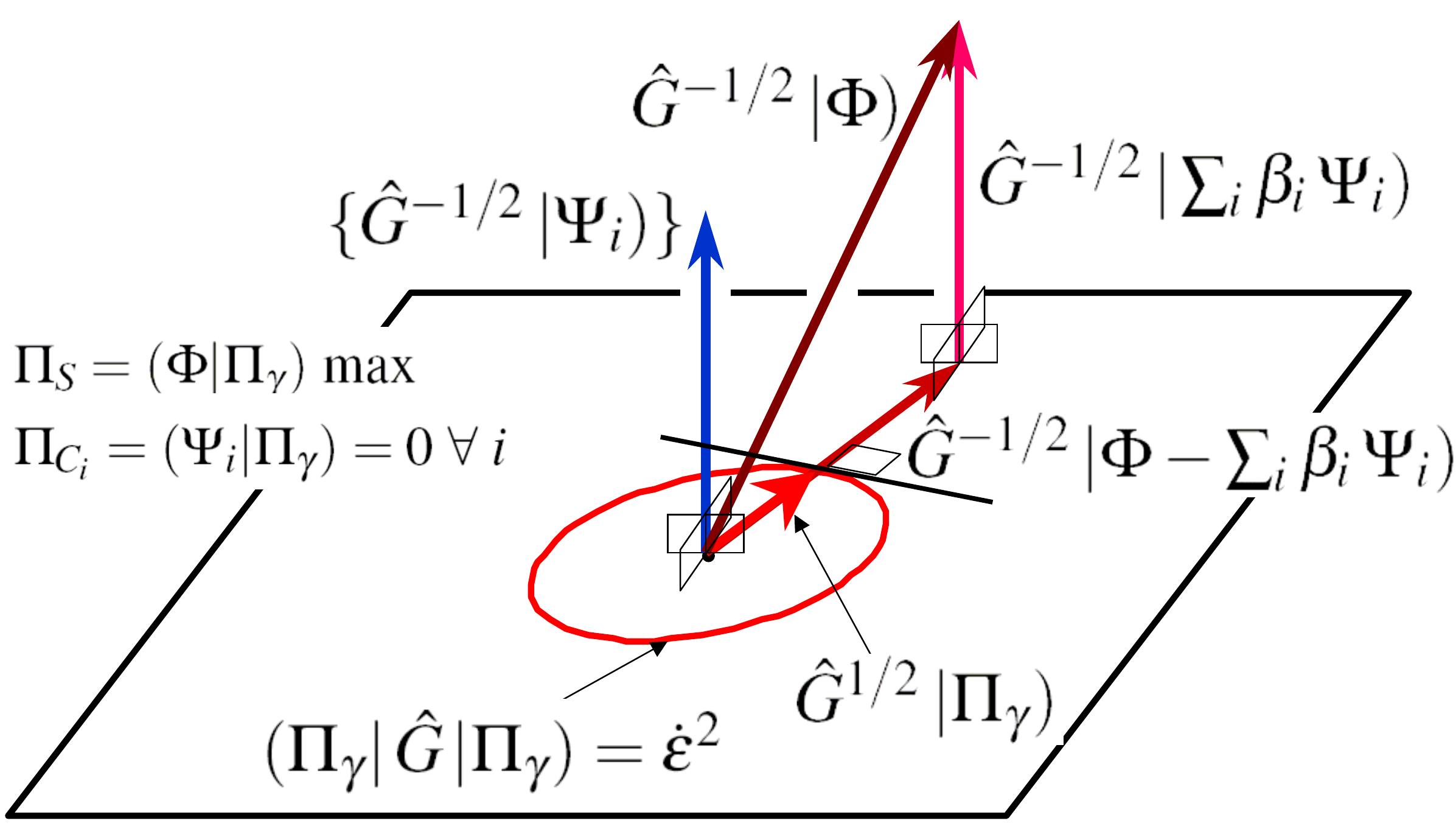}
       \caption{\label{Figure4}Pictorial representation of the SEA variational construction in the case of a non-uniform metric $\hat G$. The ellipse  represents  the  condition $(\Pi_\gamma|\,\hat G\,|\Pi_\gamma)=\dot\epsilon^2$, corresponding to the norm of vector $\hat G^{1/2}\,|\Pi_\gamma)$. This vector must be orthogonal to the $\hat G^{-1/2}\,|\Psi_i)$'s in order to satisfy the conservation constraints $\Pi_{C_i}=(\Psi_i|\Pi_\gamma)=0$. In order to  maximize the scalar product  $\Pi_{S}=(\Phi|\Pi_\gamma)=(\Phi-\sum_i \beta_i\, \Psi_i|\Pi_\gamma)$, vector $\hat G^{1/2}\,|\Pi_\gamma)$ must have the same direction as $|\Lambda)=\hat G^{-1/2}\,| \Phi-\sum_i \beta_i\, \Psi_i)$.}
   \end{figure}

Next, we consider the more general  scenario of a non-uniform metric tensor $\hat G$.
Figure \ref{Figure3} gives a pictorial representation of the linear manifold spanned by the vectors $\hat G^{-1/2}\,|\Psi_i)$ and the orthogonal projection of $\hat G^{-1/2}\,|\Phi)$ which defines the Lagrange multipliers $\beta_i$ in the case of non-uniform metric $\hat G$, where the orthogonality conditions that define the $\beta_i$'s are $(\Psi_j|\,\hat G^{-1}\,|\Phi-\sum_i \beta_i\, \Psi_i)=0$ for every $j$, which is Eq.\ (\ref{11g}).  The construction defines also the generalized affinity vector $|\Lambda)=\hat G^{-1/2}\,| \Phi-\sum_i \beta_i\, \Psi_i)$ which  is orthogonal to the linear manifold spanned by the vectors $\hat G^{-1/2}\,|\Psi_i)$'s.

Figure \ref{Figure4} gives a pictorial representation of the subspace orthogonal to the linear manifold spanned by the $\hat G^{-1/2}\,|\Psi_i)$'s that here we denote for simplicity by $\{ \hat G^{-1/2}\,|\Psi_i) \}$. The vector $\hat G^{-1/2}\,|\Phi)$ is decomposed into its component $\hat G^{-1/2}\,|\sum_i \beta_i\, \Psi_i)$ which lies in $\{ \hat G^{-1/2}\,|\Psi_i ) \}$  and its component $|\Lambda)=\hat G^{-1/2}\,|\Phi-\sum_i \beta_i\, \Psi_i)$ which lies in the orthogonal subspace.

The ellipse in Figure \ref{Figure4} represents  the more general condition $(\Pi_\gamma|\,\hat G\,|\Pi_\gamma)=\dot\epsilon^2$ corresponding in the non-uniform metric to the  prescribed rate of advancement in state space, $\dot\epsilon^2=(d\ell/dt)^2$. It is clear that the  direction of $\hat G^{1/2}\,|\Pi_\gamma)$ which maximizes the scalar product  $(\Phi-\sum_i \beta_i\, \Psi_i|\Pi_\gamma)$, is when $|\Pi_\gamma)$ is in the direction of the point of tangency between the ellipse and a line orthogonal to $| \Phi-\sum_i \beta_i\, \Psi_i)$.

The compatibility with the conservation constraints $\Pi_{C_i}=(\Psi_i|\Pi_\gamma)=0$ requires that  $\hat G^{1/2}\,|\Pi_\gamma)$ lies in subspace orthogonal to the  $\hat G^{-1/2}\,|\Psi_i)$'s. To take the SEA
 direction, the vector $\hat G^{1/2}\,|\Pi_\gamma)$ must maximize the scalar product  $(\Phi-\sum_i \beta_i\, \Psi_i|\Pi_\gamma)$, which is equal to the entropy production $\Pi_{S}=(\Phi|\Pi_\gamma)$ since $(\Psi_i|\Pi_\gamma)=0$. This clearly happens when $\hat G^{1/2}\,|\Pi_\gamma)$ has the same direction as the generalized affinity vector $|\Lambda)=\hat G^{-1/2}\,| \Phi-\sum_i \beta_i\, \Psi_i)$.

\section{Conclusions}\label{concl}

 In this paper, we reformulate with a somewhat unusual notation the essential mathematical elements of sixseveral different approaches to the description of non-equilibrium dynamics with the purpose of presenting a unified formulation which,  in all these contexts, allows to implement the local Steepest Entropy Ascent (SEA) concept whereby the dissipative, irreversible component of the time evolution the local  state  is assumed to pull the state along the path in state space which, with respect to an underlying metric, is always tangent to the direction of maximal entropy increase compatible with the local conservation constraints.

The   frameworks are: A) Statistical or Information Theoretic Models of Relaxation; B)  Small-Scale and Rarefied Gases Dynamics (i.e., kinetic models for the Boltzmann equation); C)  D) Rational Extended Thermodynamics, Macroscopic Non-Equilibrium Thermodynamics, and Chemical Kinetics; D) Mesoscopic Irreversible Thermodynamics, Continuum Mechanics with Fluctuations; E) Quantum Statistical Mechanics, Quantum Thermodynamics, Mesoscopic Non-Equilibrium Quantum Thermodynamics, and Intrinsic Quantum Thermodynamics.

   The present SEA unified formulation allows us to extend at once to all these frameworks  the SEA concept which has so far been considered only in the framework of quantum thermodynamics. However, a similar or at least closely related set of assumptions underlie the  well-known GENERIC scheme \cite{Grmela84,GENERIC1,GENERIC2} which developed independently.

In the present paper, we emphasized that in the SEA construction, a key role is played by the geometrical metric with respect to which to measure the length of a trajectory in state space. The metric tensor turns out to be directly related to the inverse of the Onsager's generalized conductivity tensor. The SEA construction can be viewed as a precisely structured implementation of the MEP principle. The formal relation between the SEA metric tensor $\hat G$ and the GENERIC dissipative tensor (usually denoted by $M$) can be established by means of a detailed technical analysis of the respective underlying mathematical landscapes.  We present such discussion in a forthcoming paper, where we discuss the analogies and differences of the SEA and GENERIC approaches and show under what conditions their descriptions of the dissipative part of the time evolution can be considered essentially equivalent.

   The  formulation discussed here constitutes a generalization of our previous SEA construction in the quantum thermodynamics framework by acknowledging the need of more structured and system dependent metrics than the uniform Fisher-Rao metric. It also constitutes a natural step towards generalizing  Mesoscopic Non-Equilibrium Quantum Thermodynamics to the far-non-equilibrium nonlinear domain.

   We  conclude that in most of the existing theories of non-equilibrium the time evolution of the local state representative can be seen to actually follow in state space the path of SEA with respect to a suitable metric connected with the generalized conductivities. This is true in the near-equilibrium limit, where in all frameworks it is possible to show that the traditional assumption of linear relaxation coincides with the SEA result.  Since the generalized conductivities represent, at least in the near-equilibrium regime, the strength of the system's reaction when pulled out of equilibrium, it appear that their inverse, i.e., the generalized resistivity tensor, represents the metric with respect to which the time evolution, at least in the near equilibrium, is SEA.

   Far from equilibrium  the resulting unified family of  SEA dynamical models is a very fundamental as well as practical starting point because it features an intrinsic  consistency with the second law of thermodynamics which follows from the nonnegativity of the local entropy production density as well as the instability of the equilibrium states that do not have the maximum local entropy density for the given local   values of the densities of the conserved properties, a general and straightforward conclusion that holds regardless of the details of the underlying metric tensor. In a variety of fields of application, the present unifying approach may prove useful in providing a new basis for effective numerical and theoretical models of irreversible, conservative relaxation towards equilibrium from far non-equilibrium states.

\section*{Acknowledgments}

\noindent The author gratefully acknowledges the
Cariplo--UniBS--MIT-MechE faculty exchange program co-sponsored by
UniBS and the CARIPLO Foundation, Italy under grant 2008-2290. This work is part of EOARD (European Office of Aerospace R\&D)  grant FA8655-11-1-3068 and italian MIUR  PRIN-2009-3JPM5Z-002. A preliminary version of this work was presented at the 12th Joint European Thermodynamics Conference, JETC2013, Brescia, Italy,  July 1-5, 2013.

\bibliographystyle{unsrt}

    \end{document}